\documentclass[creativecommons]{eptcs}
\providecommand{\event}{ICLP 2024} 

\usepackage{iftex}

\ifpdf
  \usepackage{underscore}         
  \usepackage[T1]{fontenc}        
\else
  \usepackage{breakurl}           
\fi

\usepackage{amsmath}
\usepackage{amssymb}
\usepackage{multirow}
\usepackage{amsfonts}
\usepackage{graphicx}
\usepackage{color}
\usepackage{url}
\usepackage{subcaption}
\usepackage{hyperref}
\usepackage[all]{foreign}

\definecolor{mygreen}{rgb}{0,0.6,0}
\definecolor{mygray}{rgb}{0.5,0.5,0.5}
\definecolor{mymauve}{rgb}{0.58,0,0.82}
\usepackage{listings}
\newenvironment{des}{
        \begin{list}
     {$\bullet$}
     {\topsep = 1 mm
  \labelwidth = 1 mm
  \labelsep = 1.2 mm
  \parsep = 0.2 mm
  \itemsep = \parskip
  \leftmargin = 4 mm}
        }{\end{list}}

\title{Data2Concept2Text: An Explainable Multilingual Framework for Data Analysis Narration\thanks{Research partially supported by Interdepartmental
Project on AI (Strategic Plan UniUD--22-25) and by INdAM–GNCS project CUP E53C22001930001.
F.Bertini, A.Dal~Pal\`u, and A.Formisano are members of GNCS-INdAM, Gruppo Nazionale per il Calcolo Scientifico.}}
\author{Flavio Bertini \qquad Alessandro {Dal Pal{\`u}}  \qquad Federica Zaglio
\institute{Department of Mathematical, Physical and Computer Sciences, University of Parma,\\
Parma, Italy}
\email{\{flavio.bertini|alessandro.dalpalu|federica.zaglio\}@unipr.it}
\and
Francesco Fabiano
\institute{Department of Computer Science, New Mexico State University\\
Las Cruces, New Mexico,  USA}
\email{ffabiano@nmsu.edu}
\and
Andrea Formisano
\institute{Department of Mathematics, Computer Science and Physics, University of Udine\\
Udine,  Italy}
\email{andrea.formisano@uniud.it}
}
\def\titlerunning{Data2Concept2Text: Explainable Data Analysis Narration}
\def\authorrunning{F. Bertini et al.}
\begin{document}
\maketitle

\begin{abstract}
This paper presents a complete explainable system that interprets a set of data, abstracts the underlying features and describes them in a natural language of choice. The system relies on two crucial stages: (i) identifying emerging properties from data and transforming them into abstract concepts, and (ii) converting these concepts into natural language. Despite the impressive natural language generation capabilities demonstrated by Large Language Models, their statistical nature and the intricacy of their internal mechanism still force us to employ these techniques as black boxes, forgoing trustworthiness.

Developing an explainable pipeline for data interpretation would allow facilitating its use in safety-critical environments like processing medical information and allowing non-experts and visually impaired people to access narrated information. To this end, we believe that the fields of knowledge representation and automated reasoning research could present a valid alternative. Expanding on prior research that tackled the first stage (i), we focus on the second stage, named \emph{Concept2Text}.
Being explainable, data translation is easily modeled through logic-based rules, once again emphasizing the role of declarative programming in achieving AI explainability.

This paper explores a Prolog/CLP-based rewriting system to interpret concepts---articulated in terms of classes and relations, plus common knowledge---derived from a generic ontology, generating natural language text.
Its main features include hierarchical tree rewritings, modular multilingual generation, support for equivalent variants across semantic, grammar, and lexical levels, and a transparent rule-based system.
We outline the architecture and demonstrate its flexibility through some examples capable of generating numerous diverse and equivalent rewritings based on the input concept.

\end{abstract}

\section{Introduction}
The emergence of explainable Artificial Intelligence (xAI) signifies the integration of crucial aspects within AI systems, such as transparency, ethical conduct, accountability, privacy, and fairness  \cite{ArrietaRSBTBGGM20}.
In several domains, the acceptance of AI systems depends on their ability to offer comprehensive insights into their internal workings and transparency in decision-making processes. Notably, the recent European Union AI Act aims to establish a unified legal framework to foster AI development while safeguarding public interests, such as health, safety, fundamental rights, democracy, and the environment \cite{eu-206-2021}. This legislation mandates AI systems to be sufficiently transparent, explainable, and well-documented, necessitating them to provide supporting evidence for their outputs.
These considerations are especially important when AI systems are utilized in high-risk scenarios, such as automatically describing an electrocardiogram in a medical report.
While achieving these objectives remains challenging for systems reliant on deep neural networks, it presents an opportunity for the Logic Programming community due to the inherent explainability of its products.

A simple count on papers on xAI classified by Scopus per year was analyzed by the system presented in this paper. The following is one automatically generated output:

\begin{lstlisting}[breaklines=true,basicstyle=\footnotesize\ttfamily]
From the year 2014 up to 2023 publications in explainable AI have exponentially grown in an important way (from 0 up to 1905) [excellent accuracy]; in detail, during the interval of time between the years 2014 and 2017 publications have been significantly steady (from 0 to 7) [excellent accuracy].
\end{lstlisting}

Our research focuses on transforming raw data, such as data series, into natural language descriptions within an explainable framework.
This involves interpreting and abstracting features (\emph{Data2Concept}) and subsequently translating concepts into natural language (\emph{Concept2Text}).
The first step, as shown in~\cite{DaplaCILC23,DalpaICLP23}, requires identifying user-defined patterns in raw data and representing them as high-level descriptions or \emph{concepts}. For instance, a time series showing a consistent increase in values over time can serve as a logical fact for subsequent processing, enriched with additional contextual information.
Our previous work primarily addressed the second step, i.e., Concept2Text, demonstrating a trivial natural language expression generator.
Extensions of this work have led to domain-specific applications, such as an explainable decision-making support system for analytics in academia \cite{DalpaItalIA23}.

In this work, instead, we focus on the design of a general Concept2Text pipeline whose key features include:
\begin{des}
    \item \textbf{Explainability.} Our system is grounded in Logic Programming and focuses on rewriting rules and Constraint Satisfaction Problems providing transparency at every level.
    
    \item \textbf{Modularity.} The system allows for seamless expansion to accommodate various domain-specific concepts and languages thanks to its declarative nature.
    
    \item \textbf{Tree Rewriting.} We represent concepts as trees that are manipulated progressing from the conceptual to the the syntax level generating natural language.
    
    \item \textbf{Variants.} To facilitate the generation of diverse semantic-equivalent sentences, each rewriting permits the creation of multiple versions that can be selected. Each stage of rewriting determines various levels of equivalence, ranging from conceptual and structural to grammatical and lexical.
    
    \item \textbf{Multi-Language Support.} The modular architecture of the system enables the creation of multiple language rule sets without affecting the overall structure. In this paper, this is demonstrated with examples in English and Italian.
\end{des}

The paper is structured as follows. Section~\ref{sec:relwork} provides a concise overview of the background. Section~\ref{sect:model} presents the design of the system, while Section~\ref{sec:results} demonstrates some practical results. Finally, Section~\ref{sec:conclusions} offers concluding remarks.

\section{Background}
\label{sec:relwork}

\paragraph*{Ontologies}
In information science, ontologies act as pivotal organizational frameworks, structuring knowledge within defined domains by delineating facts, properties, and their interconnections via representational elements such as classes, attributes, and relations among them \cite{staab2013handbook}.
These conceptual constructs not only explain the intricate relationships between pertinent concepts in a domain but also capture knowledge across multiple domains from the arts and sciences to cutting-edge technologies and medical sciences.

Practically, ontologies use specific languages to articulate concepts and relationships, removing the complexities of implementation.
One notable exemplar in this realm is the Web Ontology Language (OWL), which allows applications to process information and concepts in autonomy \cite{mcguinness2004owl}.
Ontologies created a profound transformation in computational reasoning, equipping machines with the capability to decipher word meanings and assemble them into intricate sentences, similar to natural language processing \cite{speer2017conceptnet}.

Within the medical domain, ontologies are a fundamental part of computational reasoning, particularly in the realms of precision medicine and explainable AI \cite{haendel2018classification}. Biomedical and health sciences extensively leverage ontologies to encapsulate a vast spectrum of knowledge, spanning diverse realms encompassing diseases \cite{schriml2019human}, gene products \cite{ashburner2000gene}, phenotypic abnormalities \cite{kohler2021human}, clinical trials \cite{smith2007obo}, vaccine information \cite{lin2012ontology}, and human anatomy \cite{noy2004pushing}.

\paragraph*{Sub-symbolic text models}
Recent advancements in the realm of natural language processing have witnessed a rise in the popularity of Large Language Models (LLMs).
These models, powered by transformer-based neural network architectures, boast an impressive capacity to manipulate, summarize, generate, and predict textual content similar to human language \cite{zhao2023survey}. Leveraging vast text corpora for training, often comprising hundreds of billions of parameters, LLMs excel in content generation within the domain of generative AI.

However, alongside their remarkable capabilities, LLMs have some fundamental limitations.
They are prone to misinterpreting instructions, generating biased content and factually incorrect information \cite{wang2023aligning}.
These drawbacks highlight a lack of control over the accuracy and consistency of the text generated, leading to concerns such as the proliferation of fake news and instances of plagiarism \cite{ji2023survey}. Such challenges align with the characterization of LLMs as black box systems, as mentioned by the EU AI Act, where comprehending and interpreting their internal mechanisms pose inherent difficulties.

\paragraph*{Symbolic text models}
The exploration of Logic Programming techniques, particularly utilizing Prolog, Answer Set Programming (ASP), and Constraint Logic Programming (CLP), for extracting concepts from raw data in alignment with xAI standards and subsequently translating them into natural language expressions, remains relatively underexplored in the literature. For instance, \cite{pereira1987grammars} proposed models of grammars and graph-based structures leveraging Prolog unification.

Conversely, there has been significant attention from the Logic Programming community towards the problem of processing natural language text as input \cite{pereira2002prolog} with Definite Clause Grammars (DCGs) playing a prominent role. DCGs, introduced in the 1980s, serve as a powerful tool for parsing both natural and artificial languages using explicit grammar rules and Prolog \cite{matsumoto1983bup}.
Notably, DCGs offer bidirectional capabilities, enabling not just parsing but also text generation from a controlled context-free grammar conforming to Backus-Naur Form rules. Additionally, they facilitate the modeling of multiple variants of the grammar tree.
While examples of formalization of language structure exist in the literature, they are not necessarily rule-based or operational \cite{den2013cambridge}. However, they still provide valuable insights into understanding and modeling language structures.

\section{Model}\label{sect:model}

\paragraph*{Overall system}

The system is composed of two modules: Data2Concept and a Concept2Text. The first one is able to identify concepts that are represented as trees of classes and relations among them. 
The second module can be fed by the first one, but in general it can process any tree concept modelization. The system can therefore describe either specific raw data or general concepts. We refer interested readers to our previous work~\cite{DaplaCILC23,DalpaICLP23} for further details on Data2Concept.

\paragraph*{Concepts}

We now focus on the design choices of the Concept2Text module. As formalized by ontologies, a general concept can be modeled by a graph where nodes are classes and (hyper-)edges are relations among them. A narration of the graph would talk about nodes (described by their names in the selected language) and would link them according to expressed relations (e.g., verbal and propositional relations). Since the graph translation into a sentence is not straightforward, let us point out a feasible sequence of manipulations that allows us to reach the goal. 

At the beginning, a general graph can not be directly translated into a narration, since nodes must be sorted into a narrative order. Moreover, not every edge can be described (especially if common knowledge is present), which requires to filter out many edges. A possible automation would require to control the summary level and the semantic information loss. In the future we plan to investigate this  aspect. From now on, we assume to work with a simple spanning tree of the original (sub)graph.

\begin{figure}[tb]
{\centering
\begin{subfigure}[ct]{0.55\textwidth}
\begin{lstlisting}[xleftmargin=1pt]
[class(student),
  [rel(attribute),attribute(plural)],
  [rel(attributive_spec),class(course)]]
\end{lstlisting}
\end{subfigure}
    \hfill
    \begin{subfigure}[c]{0.40\textwidth}
  \centerline{\includegraphics[width=\linewidth]{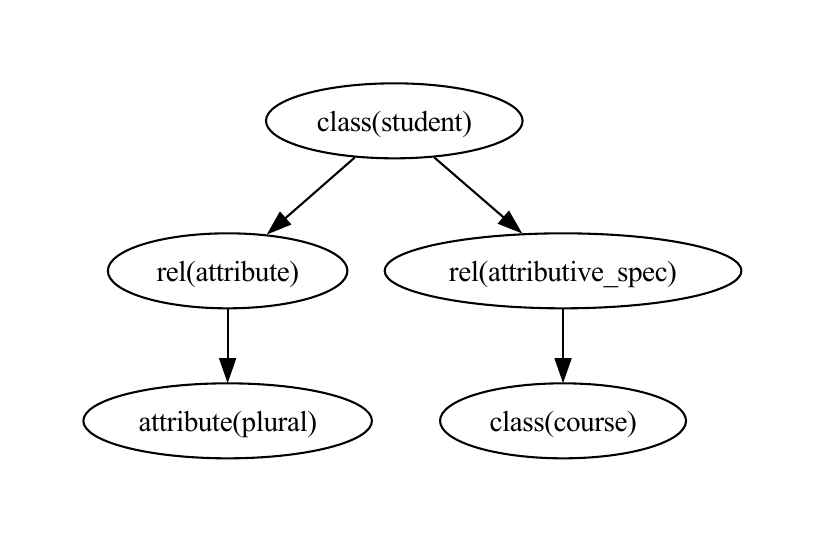}}
      \end{subfigure}
  \caption{Example of an input concept (the Prolog list and the corresponding tree).}
    \label{fig:example-input}
          }
\end{figure}

In our experimentation, trees proved to be a consistent data structure that serves the purpose of hosting concept information as well as the various translations towards a corresponding well-formed natural language expression. Therefore,  
let us illustrate a minimal example about our tree representation, in the case of the concept of \emph{students of a course}. We define two classes \emph{student} and \emph{course} by using a predicate \lstinline{class/1}, and relate them by means of \lstinline{rel/1} predicate. For example, we can specify that the class \emph{student} has the attribute \emph{plural} and specify that those students belong to a \emph{course}. Syntactically, we structure a Prolog nested list \lstinline[mathescape=true]{[Root,Child${}_1$,$\ldots$,Child${}_n$]}, where children may contain further nested lists.
Figure~\ref{fig:example-input} illustrates the encoding 
and the corresponding tree representation.

\paragraph*{Rewriting of trees} We argue that a uniform rule-based rewriting system to drive the Concept2Text process is general and modular. During rewriting, trees undergo structural (semantic) and node (syntax) modifications. 
The rewriting system must be stratified, with the idea to run a set of rewriting rules related to a specific stage until fixpoint, before starting with the next stage. This choice helps in controlling the semantics of changes and well suits for differentiating behaviours that are language dependent.

In a preliminary model we considered using DCGs, but we encountered two main limitations: DCG rules are designed to model a space of trees according to a context free grammar, rather than to control the rewriting process (when a subtree is rewritten into a new subtree). Even if adaptable for tree rewriting, the head of DCG rules does not support a general tree shape. Secondly, we need to control how alternative and equivalent rewriting options are selected, as opposed to allowing Prolog SLD resolution to explore all possibilities. 

Let us explore how tree rewriting is conceptualized (see also Section~\ref{sec:rewrite}). When making changes to a tree structure, we need to pinpoint specific conditions that trigger these modifications, typically based on the presence of certain subtrees and their relationships embedded in the larger structure. We expect a rule to encompass the lowest node that is an ancestor of all relevant subtrees, including locations that are affected by the rule rewriting (potentially elsewhere in the tree).
Each rule is responsible for constructing a new subtree that replaces the previous content hanging from that node. While some cases involve straightforward substitutions of one subtree with another, more complex scenarios can entail assembling intricate structures by combining existing components, rearranging their structure, and introducing new elements.
This level of generality enables the modeling of typical semantic equivalent rewritings as well as grammatical transformations (e.g., active vs. passive voice, word to pronoun substitution, etc.).

The Concept2Text rewriting process can be broken down into several stages, each addressing specific objectives. The overall construction of the final tree benefits from the iterative tree rewriting performed at each stage's fix points. Although the fixed-point rewriting mechanism is common across all stages, we prefer to tailor rules and introduce barriers to fix points. This approach offers several advantages, including the ability to accommodate language-independent stages alongside those requiring language-specific rules. We can also avoid to determine the complete BNF grammar of a natural language, since we can handle each stage separately. The transition from a stage to the next one requires to model rewriting rules rather than complete grammars.

We report on each stage and their purposes:
\begin{enumerate}
    \item {\bf Equivalent Concepts:} This stage rewrites classes and relations into equivalent semantic versions. The goal is to generate more specific relations that can be devised by common knowledge and to find alternative patterns to express the same meaning. The output remains a concept-based tree. This stage ensures semantic-preserving variants of the concept, leading to the furthest but still equivalent final text.
    
    \item {\bf Concept2Structure:} This stage transforms the tree of concepts and their relations into a prototype of grammar structure. It is mainly a tree shape rewriting with the addition of internal nodes that host future grammar information. It constructs the components of a sentence (noun and verbal subtrees, as well as complements). Classes and relations still retain their ontology descriptions.
    
    \item {\bf Structure2Grammar:} While maintaining the overall structure, this stage translates each class and relation into grammar lexemes and/or other simple grammatical forms.
    
    \item {\bf Coordination:} This stage ensures that subtrees are coordinated, as necessary, to match gender and number for nouns, verbs, etc.
    
    \item {\bf Inflection and Sorting:} Responsible for producing the correct inflections for nouns and adjectives, as well as conjugations for verbs. Also, it computes the correct word order for words within the same phrase through the resolution of language-dependent Constraint Satisfaction Problem (CSP).
    
    \item {\bf Syntax:} Applies local rules to consecutive words to ensure syntactic properties are met (e.g., contractions, ellipsis, etc.).
\end{enumerate}

\subsection{Implementation}\label{sec:rewrite}
Our objective is to devise rules flexible enough to handle common language properties, such as subtree swapping and restructuring. The sequence of transformations outlined in the preceding section has been realized using Prolog. Here, we provide a brief overview of the main components of this implementation.

\subsubsection{Tree rewriting}

Trees are represented as lists of the form \lstinline{[RootInfo|Children]}, where \lstinline{Children} is the list of child trees.
Each phase of the pipeline takes a tree as input and produces a list of trees as output (representing possible variants according to a rule)
obtained by applying a set of rewriting rules.
These rule sets are unique and tailored to the specific requirements of each phase. However, all rules are uniformly described by Prolog clauses defining the predicate
\begin{lstlisting}
rule(Lang,Type,Name,Tree,RewTree)
\end{lstlisting}
where
{\lstinline{Lang}} specifies the target language of the translation, which remains consistent throughout the process (currently the possible choices are English and Italian);
{\lstinline{Type}} is the specific phase of the rewriting process (i.e., \lstinline{equiv_concept}, \lstinline{concept2structure}, \lstinline{structure2grammar}, \lstinline{coordination}, \lstinline{inflection}, and \lstinline{syntax});
{\lstinline{Name}} is a unique rule ID, distinguishing a specific rewriting among those possible in the phase \lstinline{Type};
{\lstinline{Tree}} and
{\lstinline{RewTree}} are the input tree and the rewritten tree, resp.
In each phase a BFS traversal of the input tree drives rule application and is repeated until a fixpoint is reached
(i.e., no more rules of that phase are applicable).
While rewriting {\lstinline{Tree}},  an auxiliary tree \lstinline{RuleTree} is also produced.
\lstinline{RuleTree} is isomorphic in shape to \lstinline{RewTree} and describes the applied rule(s) for each node of \lstinline{RewTree}. 

It is important to note that the information gathered in \lstinline{RewTree} plays a vital role in ensuring the explainability of the approach.
\lstinline{RewTree} serves as a description of the justification for each rewrite performed.
This is achieved by recording the \lstinline{Name} argument found in the definition of the clause(s) of \lstinline{rule/5} used in the rewriting.
Currently, this information comprises rule IDs, but richer knowledge can be easily managed if needed.

In the final stage (syntax rewriting) the tree is flattened, and leaves are extracted to form a straightforward list of words comprising the sentence. 
This list of words is then rewritten, until a fixpoint is reached, using a set of rules that only inspect pairs of consecutive words.

Here are some additional details about \lstinline{rule/5}.
Each rule generates a list of trees as alternative variants.
When applying a rule, we must select one variant from this list.
We have chosen a random selection strategy that takes into account previous choices.
This approach has proven particularly effective in preventing repetitions of the same structure in different parts of the final sentence.
We keep track of the choice history for each rule using simple assertions.
In cases where the same rule is fired multiple times, we ensure that the last choice is avoided, if possible.

\subsubsection{Language independent stages}
The first stage (Equivalent Concepts), responsible for handling equivalent concepts, is language-in\-de\-pend\-ent. While classes and relations must be named according to a specific language (English in the paper), this naming convention does not affect the generation of a specific language, as names will be converted later according to language-dependent rules.

Now, let us introduce a working example to illustrate some key features contained in the stages described above. We  model the concept of an \emph{interval} that specifies the use of the class \emph{year} as its unit of measure (\emph{uom}):

\lstinline[basicstyle=\footnotesize\ttfamily]{ [class(interval),[rel(attribute),attribute(uom(class(year)))],...]}

If a common knowledge ontology is accessible, we could discover that \lstinline{is_a(year,time)} and that the class \emph{interval} can be further specified as an interval of time. Also, additional semantic knowledge about equivalences could inform the rewriting rules that an interval of time is equivalent to the class \emph{period}.

Implementing rules that trigger whenever common knowledge adds some information is straightforward. In this case, a sequence of rewritings could be:

\lstinline[basicstyle=\footnotesize\ttfamily]{ [class(interval),[rel(attributive_spec),attribute(class(time))],...]}

\lstinline[basicstyle=\footnotesize\ttfamily]{ [class(period),...]}

\noindent where \lstinline{attributive_spec} represents the specification proposition relation attached to the class \emph{time}.

Let also discuss some potential equivalences that can be drawn for an interval that deals with a range of numbers $V_1$ and $V_2$, e.g.:

\lstinline[basicstyle=\footnotesize\ttfamily]{ [class(interval),[rel(attribute),attribute(range(V1,V2))],...]}

Focusing on the treatment of the class \emph{interval}, we can propose various alternative versions. These include: (i) presenting the interval as a simple measure of a range (between\,\dots), (ii) use the interval class (e.g., the interval\,\ldots), and (iii) employing a more refined version with the addition of a relative subordinate (e.g., the interval that spans\,\ldots). Furthermore, the actual measure itself (the range between two numbers) can be expressed using different prepositions (from\,\ldots to, between\,\ldots, starting from\,\ldots up to\,\ldots). The combination of rules that rewrite different parts of the concept, sometimes even depending on one another, generates a
combinatorial explosion and produces a rich set of alternative variants already at the concept stage.

\begin{figure}[tb]
{\centering
   \begin{minipage}{0.48\textwidth}        
 \begin{subfigure}[l]{0.98\textwidth}
  {\centering
  \includegraphics[width=0.75\linewidth]{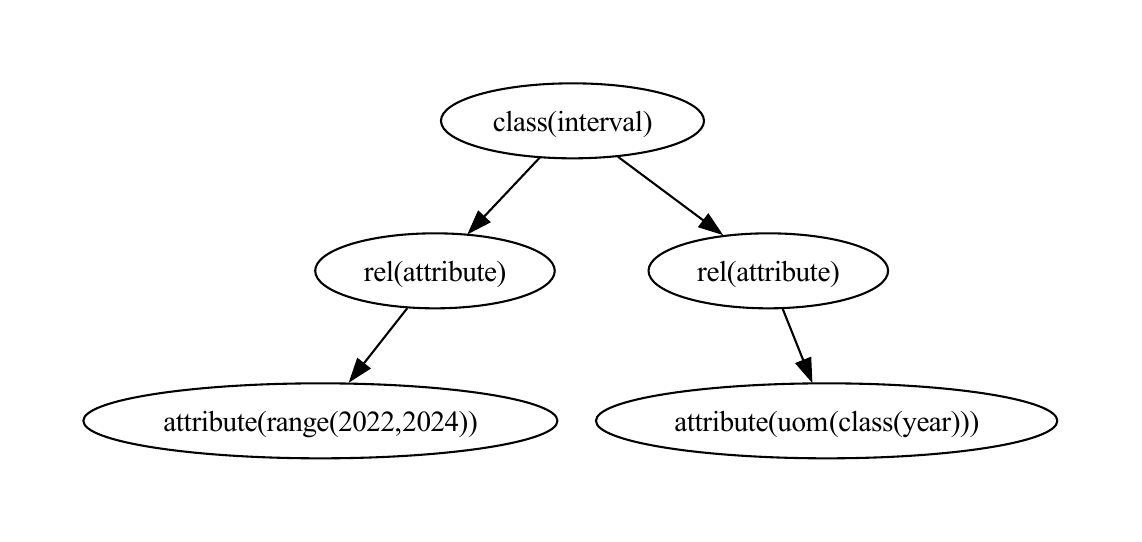}
  \caption{Input concept}
  \includegraphics[width=1.0\linewidth]{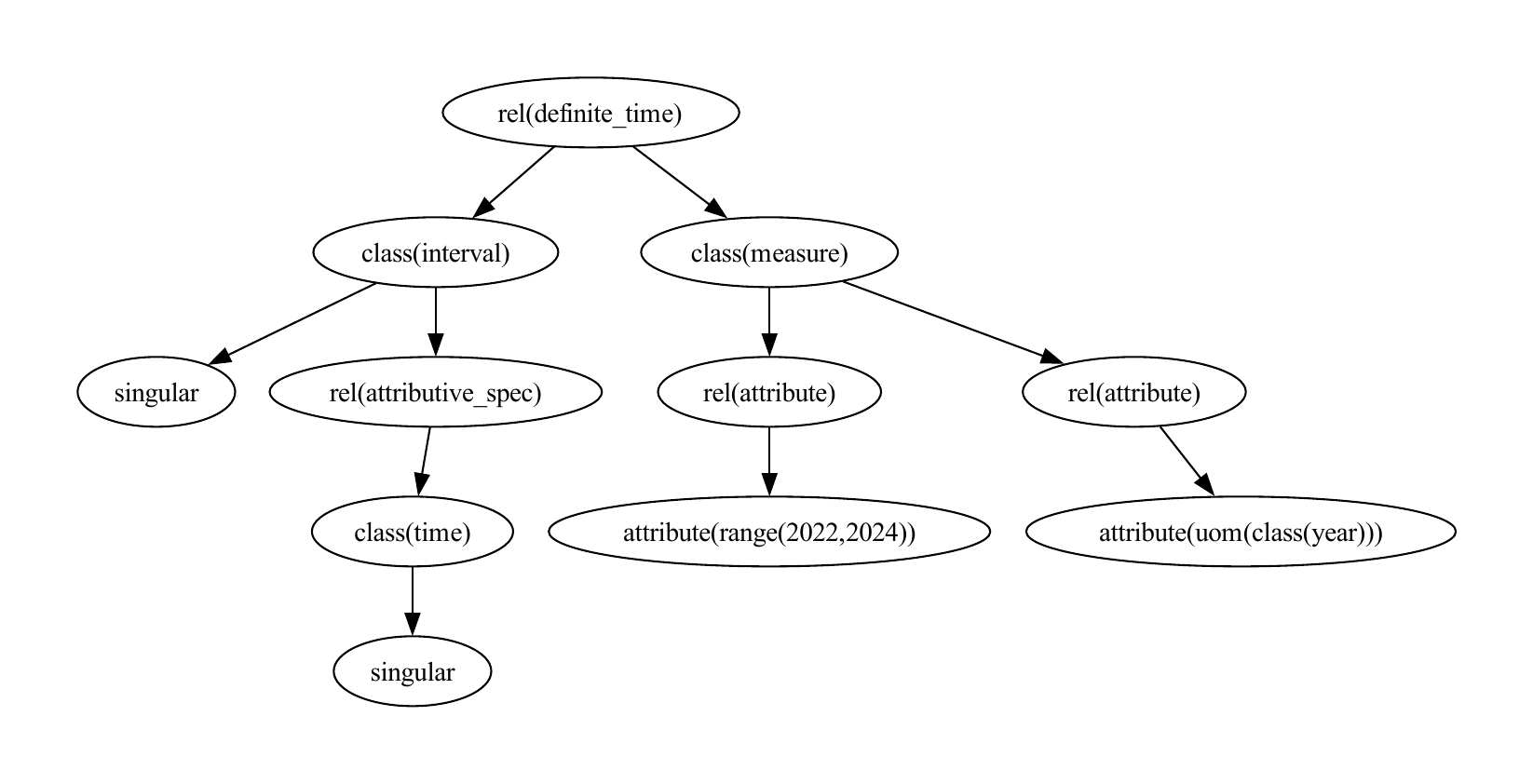}
  \caption{Rewriting of concept \emph{interval}}
   } \end{subfigure}
   \end{minipage}
   }
\hfill
\begin{subfigure}[c]{0.50\textwidth}
{\centering
  \includegraphics[width=1.0\linewidth]{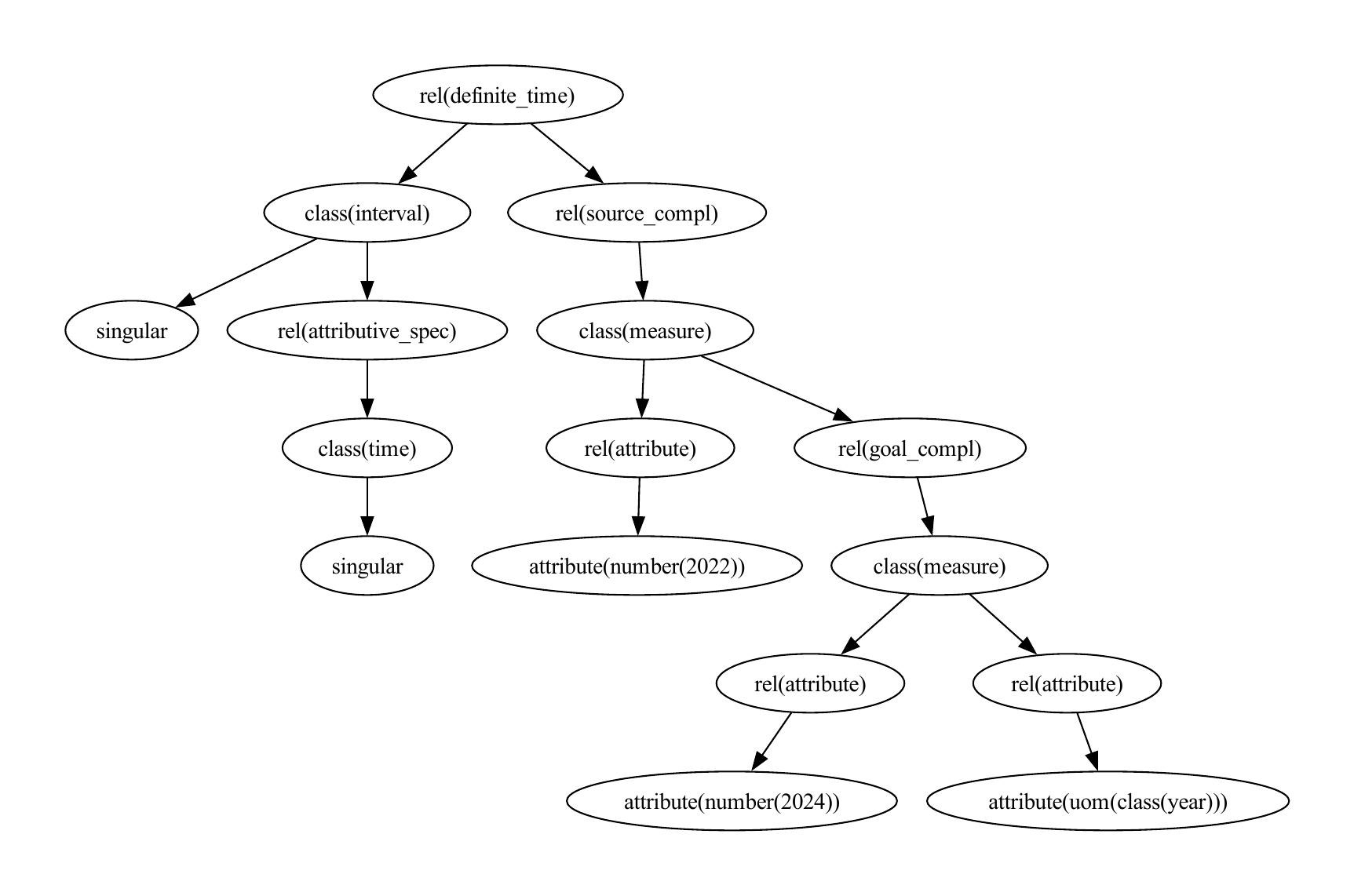}
  \caption{Rewriting of concept \emph{measure} with range}
}
\end{subfigure}
  \caption{Example of rewriting with concept equivalence.}
    \label{fig:example-equivalence}
\end{figure}

Figure~\ref{fig:example-equivalence} illustrates a simplified example demonstrating the application of the two rewriting rules described above. It is noteworthy how the classes are matched and rewritten: from (a) to (b), the class \emph{interval} is replaced with the left subtree, introducing the concept of measure; from (b) to (c), the class \emph{measure}, denoting a range, is rewritten into a nesting of two complements (source and goal) with measures of simple numeric quantities.

Let us show a simplified snippet of the first rule:
\begin{lstlisting}[numbers=left]
rule(_Lang,equiv_concept,equiv_interval, [Root|C],
      [[Root|C2], ... ] ]):- % list of equivalent trees
  % firing condition
  member([class(interval)|C1],C),
  member([rel(attribute),attribute(range(V1,V2))],C1)
  member_non_var([rel(attribute),attribute(uom(class(Uom)))],C1),
  % prepare new tree structure (depending whether isa relation is known)
  ( common_knowledge_isa(Uom,Isa),!,
    Int=[class(interval),[rel(attribute),attribute(singular)],
    [rel(attributive_spec),[class(Isa),singular]]];
    Int=[class(interval),[rel(attribute),attribute(singular)]] ),
  replace(C,[class(interval)|C1],[[rel(definite_time), Int,[class(measure)|C1]]],C2),
  ...
\end{lstlisting}
where we assume defined a predicate \lstinline{replace/4} that takes the input list, the element to be replaced, the replacement list (enclosed by an additional list, in case multiple elements need to be inserted), and returns the output list.

For the second rule applied in the example, a simplified code snippet could be:
\begin{lstlisting}[numbers=left]
rule(_Lang,equiv_class,measure_range, [Root|C],
      [[Root|C4], ... ]):-  % list of equivalent trees   
  member([class(measure)|C1],C),
  member([rel(attribute),attribute(range(N1,N2))],C1),
  ( El=[rel(attribute),attribute(uom(class(U)))],
    member(El,C1),!,Uom=[El];  % there is a UoM specified
    Uom=[] ),
  replace(C,[rel(attribute),attribute(range(N1,N2))],[[]],C2),
  replace(C2,El,[[]],C3),
  % replace subtree at measure class with single measures
  replace(C3,[class(measure)|C1],
          [[rel(source_compl),
            [class(measure),[rel(attribute),attribute(number(N1))],
             [rel(goal_compl),[class(measure),
              [rel(attribute),attribute(number(N2))]|Uom]]|Uom]]],C4),
   ...
\end{lstlisting}

The second stage (Concept2Structure) is essentially language independent. Concepts are structurally rearranged into subtrees that model grammar phrases. For instance, a class (future subject) may have a verb relation as one child, and an object may be associated with the verb as its child. This three node branch at the concept level is flattened, resulting in three ordered siblings of phrases (subject, verb, object).

It is also necessary to tag each subtree with information about their phrasal role: e.g. there are nouns, verbs, propositions and relative phrases. The rewriting is able to classify them according to relations and deductions on them.
Internal nodes are thus enriched with explicit descriptions of their subtrees, using a predicate \lstinline{info/4} that specifies the type of phrase (using standard linguistic terminology such as np, vp, pp, rp standing for noun, verbal, propositional, relative phrases), the subtype (e.g., subject, object), and the gender and number attributes that apply to the subtree.

The fourth stage (Coordination) is language independent as it is responsible for structural matching of the gender and/or number variables contained in the \lstinline{info/4} nodes. Some matches are  enforced by default (e.g., between subject and verb), but in other cases, a previous rewriting stage may have forwarded a request for an explicit coordination. For example, when creating a relative subordinate subtree, the phrase must match the gender and number to the antecedent noun. However, there is no guarantee that the noun has already been processed and the associated \lstinline{info/4} is already created. Only when that stage is over, the variables for gender and number are available. The Coordination stage can then safely find and match the correct variables, upon a request that is embedded as a service node in the subordinate tree. Coordination is enforced via unification of variables.

\subsubsection{Language dependent stages}

The overall structure allows us to focus on language-de\-pend\-ent rules for specific stages: Structure2Gram\-mar, Inflection, and Syntax. One advantage of this model is that we can easily plug in sets of rules without modifying the system.

The Structure2Grammar stage translates classes and relations into their corresponding grammar lexemes (the non-inflected roots of words) in the target language. A general application would require a complete association of synonyms for each class. While we are exploring automated tools to streamline this phase, for small domain-specific applications, manual crafting is a viable option. Synonyms are modeled as a list of lexemes in the variants of rules. The tree nodes can be replaced accordingly. Each node is encapsulated by a parent node that provides its type (e.g., noun, adjective, number, verb, preposition, etc.). This tagging allows a simple reasoning when determining the correct order of elements within a phrase. In the future we plan to find a more accurate and automated model for the choice of appropriate lexemes, based on context and common knowledge. A fluent sentence can greatly benefit from this choice.

Let us now provide some details about the inflection stage. Here, inflections for nouns, adjectives, and verbs are selected based on their gender, number and verbal tense. This is typically accomplished by consulting a dictionary that explicitly associates lexemes with words. Auxiliary verbs are generated according to the rules of the target language.

The inflection stage also arranges the words in the correct order within each phrase. Languages have various rules governing the order of words in a phrase, and attempting to cover all possible cases would be impractical due to the combinatorial explosion of possibilities.
To address this challenge, we have devised a set of rules that describe local and partial orderings among subsets of words within the phrase. By combining these partial orders, we can derive the correct total order of words. These orderings may depend on word types and specific words themselves.

We represent this network of sorting constraints as a Constraint Satisfaction Problem (CSP). While solving the CSP itself is straightforward, developing accurate order constraints requires careful tuning. Thus, the flexibility of a CSP allows to support the updates of the constraints considered.

The final stage (Syntax) addresses the enforcement of writing rules for adjacent words. Every language possesses distinct rules governing word combinations, typically controlled through local pattern matching at the word or character level. For instance, this stage handles scenarios where two words are merged into contractions or a single letter is removed/added (e.g. \emph{a increase} is converted to \emph{an increase} because of the presence of a vowel at the beginning of the second word). This stage considers the leaves of the tree output by the Inflection stage. If read according to a Depth-First Search (DFS) traversal, the list of words compose the final sentence. Internal nodes describe structural properties of sub-trees. At this stage, punctuation marks and correct spacing are handled. Moreover, internal nodes can also be exploited in case the sentence requires some markup. We experimented with HTML tagging, which can be easily incorporated into rules. 

\section{Results}
\label{sec:results}

\begin{figure}[tb]    
  \centerline{\includegraphics[width=0.98\linewidth]{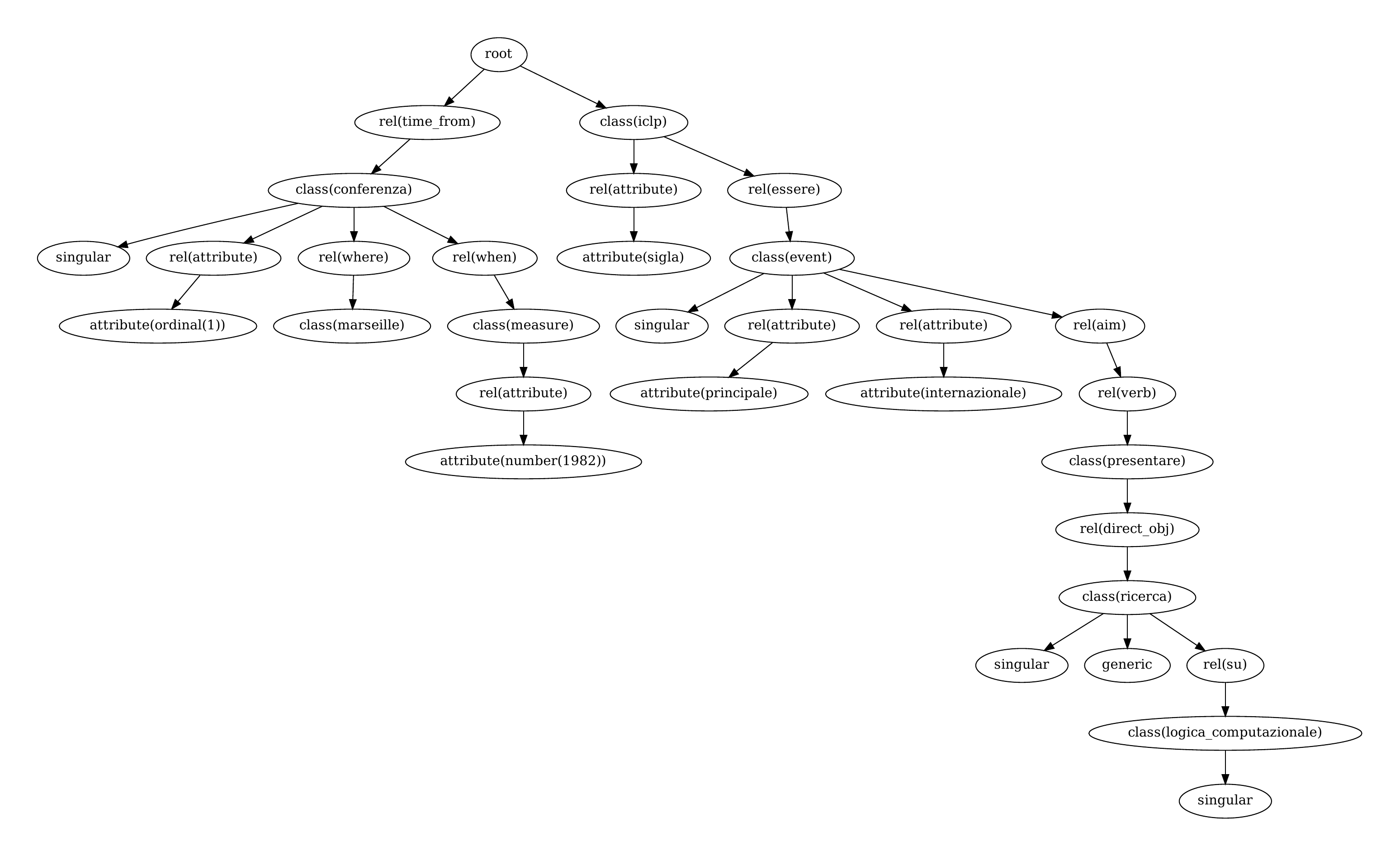}}
  \vspace{-6ex}
  \caption{Input concept for a full sentence}
  \label{fig:out0}
\end{figure}

\paragraph*{General concept narration}
We first test the potential and robustness of the Concept2Text system only. We crafted a concept that is depicted in Figure~\ref{fig:out0}. The concept describes the traditional first sentence appearing in the call for papers of ICLP2024. It can be noted that we supported generic relations of the kind \emph{where} and \emph{when} associated to the class \emph{conference}. Semantic equivalences can handle variants that involve relative subordinates, different priority in concept order and active/passive forms that greatly influence the next stages, independently on the selected language. This showcases how complex handling of relations can be performed and suggests how the underlying machinery can be adapted to many other cases, by simply changing classes and relations taken from this prototype concept.
In the supplementary material we show a sequence of tree rewritings after reaching the fixed point of each stage, starting from the input of Figure~\ref{fig:out0}. 

We locally checked each rewriting rule behaviour, along with the variants produced. This process is rather convenient while visually inspecting the rewritten trees. Clearly, the complete interaction among rules and stages grows exponentially and full checks can be performed on a single specific trace of the program. 

For the English language specialized rules set, we (non-exhaustively) collected more than 13,000 unique sentences, while for Italian we counted 3,200+ sentences. Interested readers can download and consult the list of sentences at \url{ahead-lab.unipr.it/files-for-iclp2024/}.
We manually reviewed some samples and they all appear well-formed. One advantage of the rule-based model is that errors could be quickly debugged and fixed while developing the system. Even if the variants for rules are rather limited in number (at most 4 versions per rule) and certainly they will be expanded in the future, this result shows how the combinatorics of various stages allows to enumerate a surprisingly high set of semantics preserved sentences.

It is noteworthy how the rewriting of equivalence and grammar structures is capable of providing remarkable differences, while preserving semantics perfectly, thanks to the strict transitivity of the applied equivalences. The resulting text appears natural, although some additional synonyms could be included to enrich and diversify certain words.

We also tested some free LLMs available online (i.e., ChatGPT 3.5, Gemini). We asked to generate 10 instances of a sentence that strictly preserved the semantics intended by the Prolog list as in Figure~\ref{fig:out0}. Even if in general the results were rather accurate, in some cases some attributes were skipped and/or some verb choices were not perfectly compatible with the context. It was also difficult to force the complete adherence with the original input via prompting. Those are a set of secondary issues, since the tested methodology does not comply with the major requirement of explainability.

Comparing the control obtained from a well-crafted prompt to the LLM, our pipeline ensures that the overall semantic integrity is maintained consistently from input to the final sentence throughout each step of rewriting. In contrast, LLMs may introduce arbitrary choices or hallucinations, affecting both semantic and syntactic levels. Additionally, we observed that it is challenging to enforce the use of all provided attributes, as LLMs tend to interpret the intended semantics with limited control over the degree of summarization and the relative importance of each attribute.

\paragraph*{Data2Concept narration}

Our Data2Concept system (see introduction) analyzes data series and outputs concepts in the form of trees that contain data properties as well as confidence about accuracy and relevance of the findings. The Concept2Text system can be attached and the full pipeline can be tested. We report here on a simple data analysis run on the number of papers about explainable AI indexed by Scopus each year from 2014 to 2023. The data series (y values) is $[0,2,0,7,84,217,428,816,1266,1905]$  and the contextual classes are about the x axis (year), the x axis list of values (numbers between 2014 and 2023), the y axis class (papers) and the overall class of the data series (publications on xAI). The accuracy of the extracted concepts are computed on a integer value on the range $0\dots 100$, and they are mapped into a judgment scale of adjectives. Table~\ref{tab:output} shows two sentences produced by the pipeline for both English and Italian. We can note that the adherence to the original series is very high, as stated by the accuracy feedback.

\begin{table}[ht]
{\small
\caption{Output examples. 1--2 for English and 3--4 for Italian.}\label{tab:output}
\begin{tabular}{c}
\hline
\begin{minipage}{0.98\textwidth}
~\\1. From the year 2014 up to 2023 publications in explainable AI have exponentially grown in an important way (from 0 up to 1905) [excellent accuracy]; in detail, during the interval of time between the years 2014 and 2017 publications have been significantly steady (from 0 to 7) [excellent accuracy].\\
\end{minipage}\\
\hline
\begin{minipage}{0.98\textwidth}
~\\2. There has been an important exponential growth of publications on explainable AI (from 0 up to 1905) during the interval of time that has spanned starting from the year 2014 up to 2023 [excellent accuracy]; specifically, between the years 2014 and 2017 publications have shown themselves to be significantly constant (from 0 up to 7) [excellent accuracy]. \\
\end{minipage}\\
\hline
\begin{minipage}{0.98\textwidth}
~\\3. C'è stato un incremento esponenziale importante di pubblicazioni sulla IA spiegabile (da 0 a 1905) dagli anni 2014 ai 2023 [accuratezza ottima]; in dettaglio, dall'anno 2014 e durante i 3 anni successivi le pubblicazioni sono state decisamente stabili (da 0 fino a 7) [accuratezza ottima].\\
\end{minipage}\\
\hline
\begin{minipage}{0.98\textwidth}
~\\4. Nell'intervallo di tempo dagli anni 2014 ai 2023 le pubblicazioni sulla IA spiegabile sono aumentate esponenzialmente in modo importante (a partire da 0 fino a 1905) [accuratezza ottima]; in dettaglio, nel periodo dall'anno 2014 fino al 2017 i lavori sono stati decisamente costanti (da 0 a 7) [accuratezza ottima].\\
\end{minipage}\\
\hline
\end{tabular}}\end{table}

\section{Conclusions}
\label{sec:conclusions}

The paper presented an explainable methodology to rewrite a concept in well-formed natural languages. The system adopts a multi-level rewriting procedure that can produce semantic, grammar and lexical variants that are aware of the context. Common knowledge can be used to better adapt to specific contexts. The system is  modular, since it allows for easy adaptation to various domain-specific applications and output languages. Moreover, the system adheres to explainable AI standards by offering transparency and verifiability. 
We can conclude that the Data2Concept2Text complete system is effective in  modeling general concepts and to translate them into multiple languages with the same architecture. As applications, it can handle both raw data series and narration of general concepts from ontologies.
 
This work opens different lines of research to be further explored. Adapting different rules for handling grammar and syntax from different languages requires some human time, since this kind of formalization is often fuzzy and requires language experts.
Developing a comprehensive rule set for accurately translating classes into suitable grammar synonyms is a complex task, as the most appropriate choices heavily depend  on context. We aim to devise automatic methods to retrieve such preferences and stylistic usages.

Variants of rules can be classified according to verbosity and style. This information can help to match preferences about properties of the final text to be produced.

We also plan to investigate the handling of general concept graphs, rather than our tree-like set of relations on concepts. Converting this graph into a spanning tree, or alternatively, synthesizing the information in a controlled manner, could facilitate the creation of guided concept summaries.

The entire pipeline is versatile and applicable across various domains, particularly in scenarios where reports are generated based on data analytics. We intend to extend this methodology to automated analysis of ECGs and other medical data, financial data, and more broadly, to produce trustworthy Business Intelligence (explainable automated reporting).

\bibliographystyle{eptcs}
\bibliography{DaDoF}
\subsection*{Appendix}
\begin{figure}[ht]
\centerline{\includegraphics[width=0.7\linewidth]{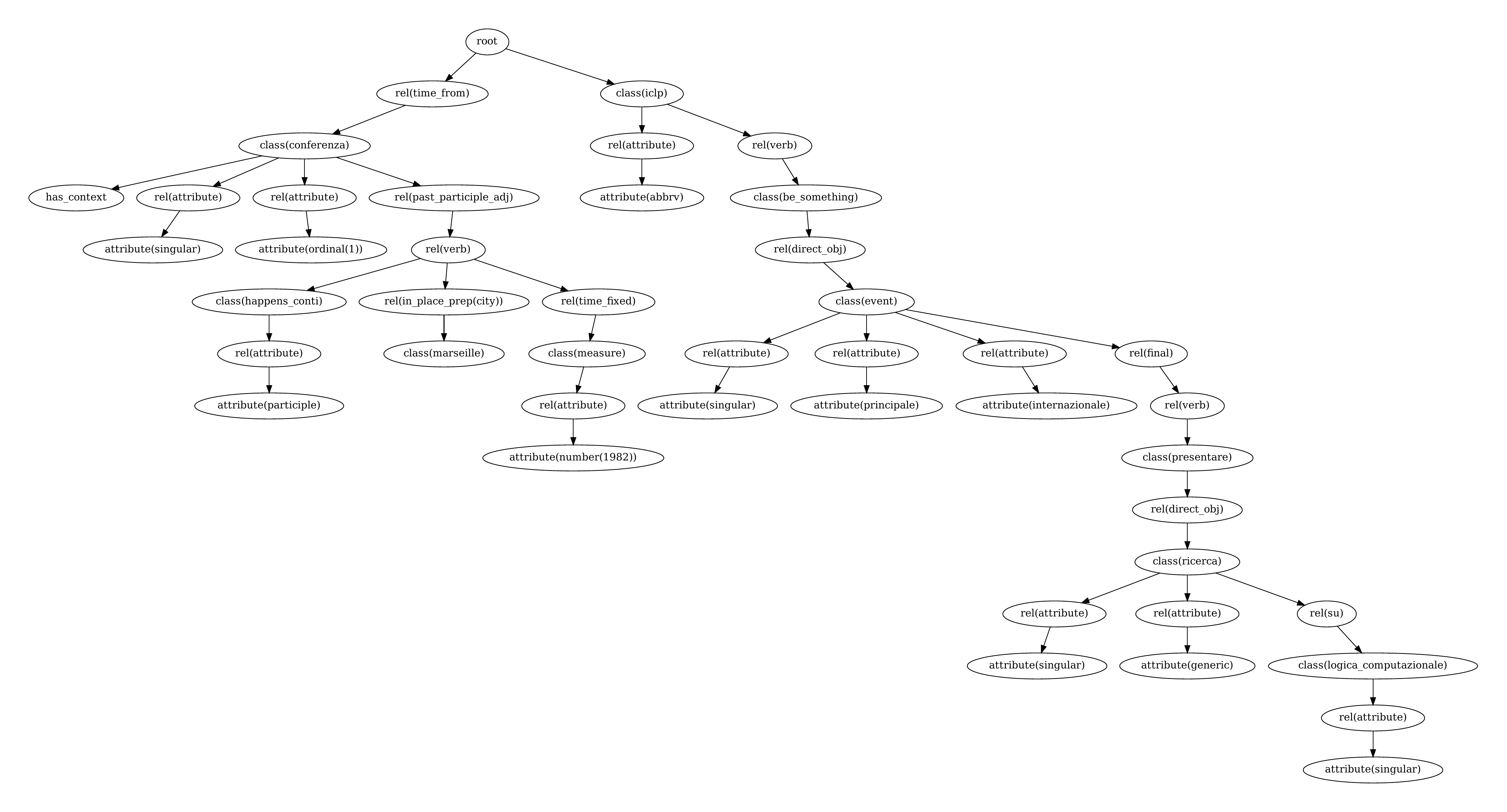}\vspace*{-12ex}}
\centerline{\includegraphics[width=0.9\linewidth]{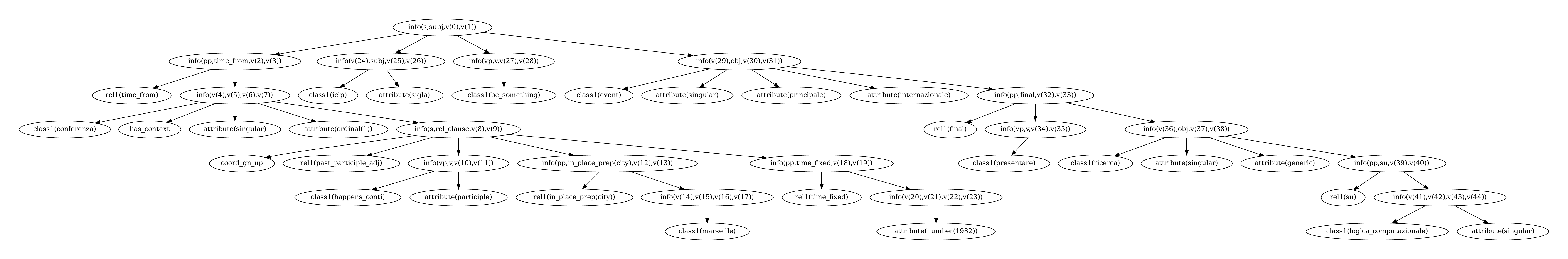}}
\centerline{\includegraphics[width=0.9\linewidth]{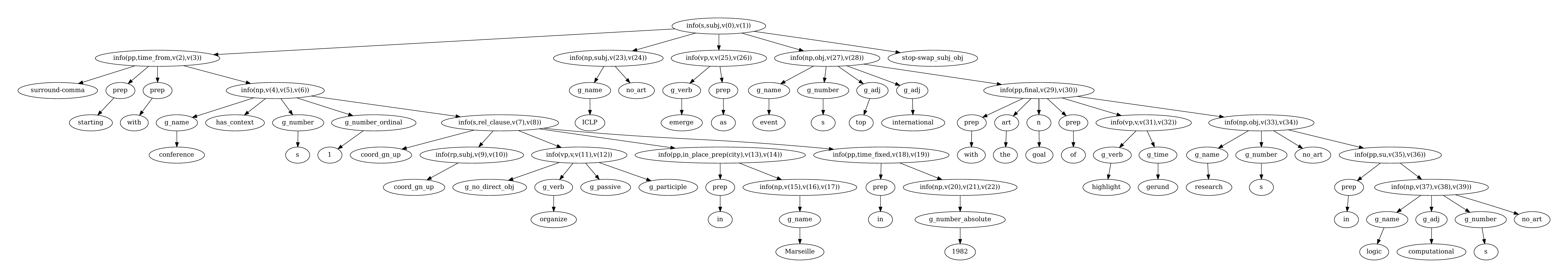}}
\centerline{\includegraphics[width=0.9\linewidth]{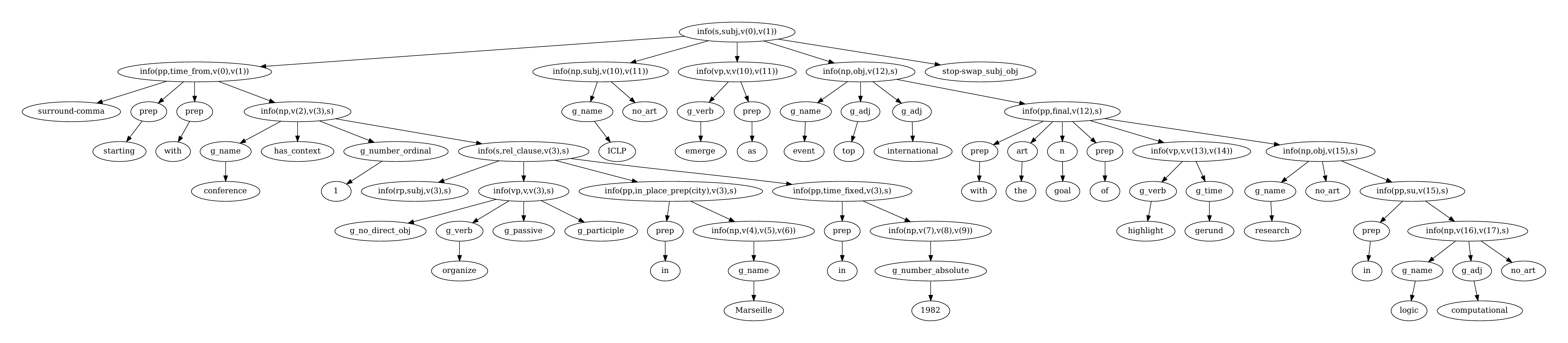}}
\centerline{\includegraphics[width=0.9\linewidth]{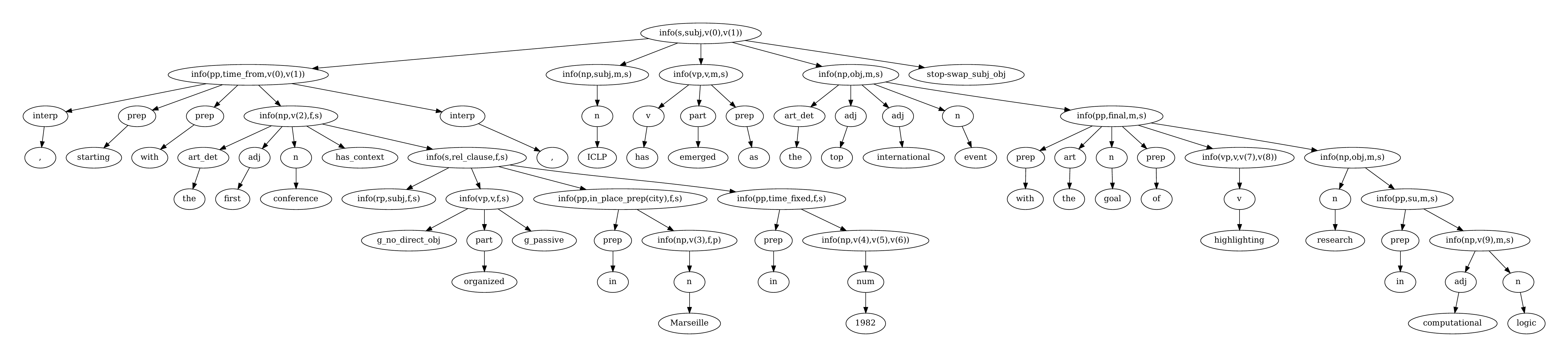}}
\caption{Computed stages for the call for papers concept of Figure~\ref{fig:out0}. From top to bottom: Concept Equivalence, Concept2Structure, Structure2Grammar, Coordination, Inflection}
\label{fig:result1}
\end{figure}
\end{document}